Numbers of words    4699

Estimated pages: 4

Tingting Jia[1], Hideo Kimura[1], Zhenxiang Cheng[2], Hongyang Zhao[3]

[1] National Institute for Materials Science, 1-2-1 Sengen, Tsukuba, Ibaraki 305-0047,

Japan

[2] Institute for Superconducting & Electronic Materials, University of Wollongong,

Innovation Campus, North Wollongong, NSW 2500, Australia

[3] Department of Materials Science and Engineering, Wuhan Institute of Technology,

Wuhan 430074, China

**Switching of both local ferroelectric and magnetic domains in multiferroic $Bi_{0.9}La_{0.1}FeO_3$ thin film by mechanical force**

Cross-coupling of ordering parameters in multiferroic materials by multiple external stimuli other than electric field and magnetic field is highly desirable from both practical application and fundamental study points of view. Recently, mechanical force has attracted great attention in switching of ferroic ordering parameters via electro-elastic coupling in ferroelectric materials. In this work, mechanical force induced polarization and magnetization switching were investigated in a polycrystalline multiferroic $Bi_{0.9}La_{0.1}FeO_3$ thin film using a scanning probe microscopy system. The piezoresponse force microscopy and magnetic force microscopy responses suggest that both the ferroelectric domains and the magnetic domains in $Bi_{0.9}La_{0.1}FeO_3$ film could be switched by mechanical force as well as electric field. High strain gradient created by mechanical force is demonstrated as able to induce ferroelastic switching and thus induce both ferroelectric dipole and magnetic spin flipping in our thin film, as a consequence of electro-elastic coupling and magneto-electric coupling. The demonstration of mechanical force control of both the ferroelectric and the magnetic domains at room temperature provides a new freedom for manipulation of multiferroics and could result in devices with novel functionalities.

Multiferroic materials combine two or three types of ferroic ordering simultaneously, such as ferroelectric, ferromagnetic, and ferroelastic orderings, and therefore, they present tremendous opportunities to investigate the coupling between these properties, which enables the dynamic manipulation of one ordering parameter by another for a broad range of potential applications in sensing, actuation, memory, etc.[1, 2] Numerous efforts have been made to investigate the coupling phenomena (Figure 1), especially the coupling between electric and magnetic orderings that would account for the magneto-electric (ME) effect in multiferroic materials.[3] $BiFeO_3$ (BFO) has been one of the most popular and well-studied materials in the multiferroic research field, not only because of its simultaneous ferroelectricity and antiferromagnetism at room temperature, but also due to the intrinsic ME coupling at room temperature.[4, 5, 6] There are no reports, however, on the elasto-electric and elasto-magnetic coupling effects of $BiFeO_3$ materials compared with the extensively studied ME coupling.

Flexoelectricity in ferroelectric thin films is considered to be the result of elasto-electric coupling, which is attributed to an electric field generated by a strain gradient. This phenomenon was proposed in 1964[7] and has been studied ever since. Nevertheless, most of the investigations on flexoelectricity in bulk solid ferroelectrics show that the effect of flexoelectricity is negligible. Recently, people have started to focus on the flexoelectric properties of thin films, because such effects becomes significant or even dominant over the piezoelectric effect on the nanoscale.[8, 9, 10] Due to the existence of the ME coupling effect between ferroelectric and magnetic orderings in multiferroic materials, flexoelectric effects where there is strain-gradient-associated electric polarization in multiferroic materials, especially in nanosized multiferroic materials, would lead to significant coupling between the strain gradient and the magnetic ordering. Although the mechanical-force-induced ferroelectric phase transition in multiferroic $BiFeO_3$ thin film has been studied experimentally,[11] to the best of our knowledge, mechanical force induced switching of both magnetic and ferroelectric domains has never been realized in any multiferroic material. Scanning probe microscopy (SPM) based techniques have emerged as powerful characterization methods to probe functional materials on the nanoscale. A spectacular example is piezoresponse force microscopy (PFM), which has opened a pathway for not only probing the polarization and electromechanical response on the nanoscale level, but also applying mechanical force locally on the material. One experimental approach to induce flexoelectricity in nanosized ferroelectric materials is using a PFM tip to apply mechanical force on the material, [12] which will generate a stress field with a large field gradient. The stress field divergence persists throughout the film, and the strain gradient drives atomic displacement, which builds in an internal electric field and results in a polarization reversal. Therefore, we can use PFM to apply nanoscale mechanical force on nanomaterials through a probe tip, and scan the ferroelectric response and magnetic phase by changing the working mode from PFM to magnetic force microscopy (MFM), so as to study both ferroelectric and magnetic domain switching induced by mechanical force in our multiferroic thin film. Generally, mechanical strain can generate electric polarization ($P_i$) in a polar material through the following equation[13]:

$$P_i = d_{ijk}\sigma_{jk} + \mu_{ijkl}\frac{\partial \varepsilon_{jk}}{\partial x_l},$$

(1)

The first term is the dielectric response, where $d_{ijk}$ is the piezoelectric coefficient, and $\sigma_{jk}$ represents the stress uniformly distributed across the sample; the second term refers to the flexoelectric response, where $\mu_{ijkl}$ is the flexoelectric coefficient, and ($\partial \varepsilon_{jk} / \partial x_l$) is the strain gradient. During probing with high loading force on the PFM tip, widespread stress is generated in the film: the stress distribution is uniform along the in-plane (IP) direction, but varies along the out-of-plane (OP) direction and induces a strain gradient which can generate electric polarization through the flexoelectric effect. It is obvious that Eq. (1) only describes the electric polarization and its mechanical origin because the previous research was limited to only the ferroelectrics, so how the strain gradient operates on the magnetic ordering in multiferroics is still unknown. It has been reported that giant magneto-elastic coupling can lead to atomic displacements and thus give rise to strong ME coupling in rare earth magnets, [14] which may lay a solid foundation for the possible experimental switching of magnetic domains by mechanical force in multiferroic materials.

In this work, we deposited $Bi_{0.9}La_{0.1}FeO_3$ (BLFO) thin film on $Pt/TiO_2/SiO_2/Si$ substrate by pulsed laser deposition (PLD) (crystal structure characterization and component analysis of the film are shown in Figure S1), which is an optimized composition of $BiFeO_3$ with better electric performance in terms of reduction in current leakage, for our tests of mechanical switching of both ferroelectric and magnetic domains. A large remanent polarization ($2P_r$) of 170.7 $\mu C/cm^2$ and a coercive field ($E_c$) of about 303.6 kV/cm at a maximum applied field of 500 kV/cm were observed. (See details in Supporting Information Figure S2) We have visualized the switching of magnetic domains in the as-deposited BLFO thin film by applying mechanical force on the probe tip using a Nanocute SPM system. Meanwhile, we switched magnetic domains as well as ferroelectric domains using an electrical field. Therefore both ferroelectric and magnetic domain switching were successfully demonstrated by mechanical force as well as by electric field, revealing the complex coupling among polarization, magnetization, and strain in multiferroic BLFO thin film.

Figure 2 shows the ferroelectric and magnetic characterization of the as-deposited BLFO thin film by PFM and magnetic force microscopy (MFM). Figure 2(a) presents the topographic image. Figure 2(b, c) shows the corresponding vertical piezoresponse force microscope (VPFM) image and the lateral piezoresponse force microscope (LPFM) image after alternatively applying dc bias of ±15 V with a box-in-box pattern, respectively. The VPFM shows the out-of-plane (OP) response, while the LPFM represents the in-plane (IP) components. The dark and bright contrast in the VPFM image represents upward and downward polarizations [Figure 2(f)], respectively. The observed domain reversal when the applied dc bias switches from positive to negative demonstrates the ferroelectricity of our thin film. The domain switching observed in VPFM mode as the electric field changes from positive to negative comes from the OP component of the polarization of the BLFO thin film. The LPFM image shows the IP component, which can be inferred from the shear forces experienced by the tip. The domain switching behavior in BLFO thin film through the VPFM and LPFM modes is illustrated in Figure 2(e). On applying a bias between the surface and bottom electrodes of the BLFO thin film, an inhomogeneous OP electric field is generated, and the ferroelectric polarization of BLFO can point along eight possible <111> directions.[15] Three types of switching are possible: $71^{\circ}$, $109^{\circ}$, and $180^{\circ}$, classified by the angle of rotation of the polarization across the domain wall.[16] The $71^{\circ}$ switching of the spontaneous polarization, due to the reversal of only the polarization component parallel to the applied field, usually shows a striped domain structure due to the 4-variant domain walls. [17] The $109^o$ domain walls, however, show a mosaic or disordered domain structure.[18] The $180^o$ rotation of the polarization may occur through the combination of $71^o$ and $109^o$ rotations driven by the action of the electric field and the elastic interactions with the surrounding unswitched $BiFeO_3$ matrix. Due to the fact that our BLFO film is polycrystalline with non-epitaxial growth, the observed switching behavior by PFM is a collective behavior of all the switchable dipoles under the electric field perpendicular to the surface, and the VPFM and LPFM images result from the projections of these dipoles along different directions. Owing to the relative geometry between the applied field and the polarization of the as-deposited BLFO thin film, the electric field along the OP direction is effectively applied to switch the vertical component, but it does not have a direct driving force to switch the IP component of the polarization, and thus, the corresponding box-in-box pattern in the VPFM image is clearer than in the LPFM image. The true orientation of the polarization can be deconvoluted only if all the components of the pizoelectric tensor are known.[19] Because the orientations of the individual grains are random for our polycrystalline BLFO thin film, quantitative analysis of the switching of these domains is difficult. Magnetic domain switching was observed in MFM mode after the electric poling, as shown in Figure 2(d), indicating a clear magnetoelectric (ME) coupling in the as-deposited BLFO thin film, which is consistent with Chu et al.'s work, which has proved the electric field control of local ferromagnetism in a heterostructure of $BiFeO_3/Co_{0.9}Fe_{0.1}$ through the interaction at the interface.[20] The IP $71^o$ domain switching is expected to drive an in-plane rotation of the ferromagnetic plane.[18, 21] The OP electric field can also drive the OP $71^o$, $109^o$, and $180^o$ ferroelectric switching, but only the OP 109$^{\circ}$ one could rotate the antiferromagnetic axis on a (001) BFO thin film.[22] Note that the PFM scans are in contact mode, while the MFM is in lift mode, and thus, the response signals are completely different from each other. A schematic representation of PFM and MFM is presented in Figure 2(f). During the MFM measurements, because the short-range interaction is negligible in the scanning distance, there are three interactions possible between the magnetic tip and the sample: the long-range electrostatic force ($F_e$), the medium range van der Waals interactions ($F_{vdW}$), and the magnetic force ($F_m$), so the total force ($F_t$) could be written as[23]: $F_t = F_{vdW} + F_e + F_m$. Assuming that the $F_{vdW}$ interaction dominates at distances below 1 nm, at the tip-sample distance of 10 nm during MFM imaging in the present work, the $F_{vdW}$ can be neglected. The $F_e$ could be compensated by applying the correct compensation voltage.[23] So, only the magnetic interaction is recorded in the trace mode of the second scan following the topographical imaging in the first scan [Figure 2(f)].

To investigate the effects of both electric field and mechanical force on the electric polarization and magnetization in BLFO thin film, we switched the sample both by electric field and by mechanical force, and observed the switching area in PFM mode and MFM mode. Figure 3(a) presents a topographic image of the BLFO thin film, where a similar smooth surface to Figure 2(a) was observed. We firstly poled the thin film with +15 V and -15 V dc bias in turn in the outer square boxes with areas of 8×8 $\mu m^2$ and 6×6 $\mu m^2$, respectively; and then we poled the thin film with +15 V in the 4×4 $\mu m^2$ box again, and applied a mechanical force of 800 nN without applying any dc bias in the small square 2×2 $\mu m^2$ box in the centre [Figure 3(b)]. A clear box-in-box pattern is observed, corresponding to the upward and downward ferroelectric domains switched by electric field and/or mechanical force. Subsequently, we scanned the same area in MFM mode using a magnetic tip without applying dc bias, and a similar image to the PFM box-in-box pattern was observed [Figure 3(c)], indicating that the magnetic domains in the as-deposited BLFO thin film were simultaneously switched by electric field or mechanical force. The scan rate was 5 kHz, corresponding to a tip-surface contact time of 500 s/nm, so the scan is comparable to locally applying a force, and the mechanical force indicated by the PFM tip was effectively applied on the sample. The mechanically switched area is responsive to the electrical bias and can be switched, depending upon the polarity of the bias, while the magnetic domains are also switched along with the ferroelectric domain switching.

To further investigate the effects of mechanical force on the electric polarization and magnetization, we observed the domain switching while increasing the loading force on the sample. The PFM response difference ($\Delta mV$) of the *Acos* image between the electrically switched area and the force switched area (which is equivalent to a switching polarization of the mechanical force) as a function of the loading force is shown in Figure 3(d). Initially, the switching polarization difference increases as the loading force increases, and then, at a loading force of 600−700 nN, the switching polarization difference reaches a maximum. On further increasing the loading force, the switching polarization difference will decrease. This behavior is analogous to the polarization-reversal process by applying an electrical voltage in conventional PFM, in which the electromechanical response passes through zero during switching,[9] so that there is a minimum of the amplitude when the loading force is increasing, while the electrically switched polarization is fixed when applying the same voltage, so the switching difference shows a maximum while the loading force is increasing. On applying an external mechanical force, a widespread strain gradient is generated, so a flexoelectric field ($E_i$) is generated in the film, and the field can be regarded as an analogue to the electric field, which can modify the free energy profile asymmetrically in contrast to homogeneous stress.[24] (Figure S3) When the flexoelectric field $E_i$ is large enough to overcome the energy barrier, the polarization flips. Figure 3(e) presents a similar trend in the mechanical-magnetic phase reversal ($\Delta^o$), interpreted by magnetic phase in MFM images as a function of loading force, which shows that the magnetic switching reaches a maximum at 600 nN, indicating an inherent relationship between the magnetic domain switching and the ferroelectric domain switching in the BLFO thin film by mechanical force. As we mentioned above, when applying the loading force from the tip onto the BLFO thin film, the mechanical force induces a strain gradient, resulting in an internal flexoelectric field ($E_i$). The flexoelectric field will lead to a symmetry breaking and thus not only reverses the polarization, but also modulates the antiferromagnetic vectors in the film. The orientation of the antiferromagnetic magnetization is coupled to the intrinsic ferroelastic strain state of the system and is always perpendicular to the ferroelectric polarization in $BiFeO_3$.[6] When the force is acting on the domains, the polarization of the ferroelectric domains is switched, and the orientation of the easy magnetization plane is changed. Therefore, the mechanical-force-induced ferroelectric switching would lead to a reorientation of the antiferromagnetic order, and the magnetic domains could be switched. (Schematic illustration is shown in Figure S4.)

We realized that the mechanical force could switch the ferroelectric domains and magnetic domains only when the domains were initially switched by electric field.

Applying a mechanical force directly on the virgin surface of the as-deposited thin film seems to have little effect on the magnetic domains. It is possible that when the mechanical force was applied on the fresh polycrystalline film with randomly arranged domains, although a strain gradient was generated which would switch individual domains, the rotation of the chirality of the domains did not follow a particular direction. Thus, there is still no obvious contrast observed in the MFM image due to the random distribution of magnetic domains after mechanical switching without an initial electrical switching. We consider that this is due to the reality that there is no "vector" relationship between the polarization and the extrinsic stress induced by mechanical force, in drastic contrast to the "vector" relationship between the polarization/magnetization and the electric/magnetic fields. Once the domain has been switched with the OP polarization component pointing upward, due to the ME coupling in the BLFO thin film, the magnetic domains are also arranged in a certain "order", so it is easier to switch back the domains by a strain gradient because the energy barrier along the OP direction is lowered by the electric field. Furthermore, the flexoelectric effect due to the loading force is similar to that created by an electric field which changes the free energy profile asymmetrically, which would also take account of the factors for switching the magnetic domains. The contribution for the shear component is, however, an order of magnitude smaller than for the OP component. Mechanical stress or strain can generate electric polarization both from the piezoelectric response and from the flexoelectric response in ferroelectric materials. Considering that the piezoelectric effect induces a symmetrical contribution to the free energy,[25] we could not exclude the piezoelectric effect in our film. Both the flexoelectric effect and the piezoelectric effect should be taken into account for switching the ferroelectric and magnetic domains.

In the process of scanning the film surface with a PFM tip, charge injected into the film from the charged tip will accumulate at the film surface. In addition, there is also screening charge accumulated at the polar surface. All these charges at the film surface will give a false response in the PFM phase image and cover/mask the intrinsic polarization response. Therefore, we carried out further work to eliminate the screening charge effect. A VPFM lithographic image is shown in Figure 4(a), and the switching process is the same as that mentioned above. Next, the scanning tip and the bottom electrode were short-circuited, and no contrast was observed during the short-circuit process, as shown in Figure 4(b), with reference to the screening of the film surface charges; finally, we read the area again without the short-circuit between the top and bottom electrodes, and found that the lithographic pattern could still be obtained [Figure 4(c)]. The pattern, which showed strong contrast caused by the polarity of the surface after applying the electric fields and loading forces, was almost the same as was scanned before the short-circuit, indicating that the observed phase contrast is intrinsically from the switching of ferroelectric domains.[26] Figure 4(d) presents a schematic diagram of the measurement.

**Conclusion**

In summary, we have demonstrated that both ferroelectric and magnetic domains in our $Bi_{0.9}La_{0.1}FeO_3$ thin film on $Pt/TiO_2/SiO_2/Si$ substrate could be switched by loading mechanical force and applying electric field. We have proved that flexoelectricity caused by the strain gradient when an external loading force is applied can emerge as a practical means to control both ferroelectric and magnetic domains in BLFO thin film. This result provides a pathway towards understanding the ferroelectric-magnetic-elastic coupling mechanism in multiferroic materials. Furthermore, it will enrich our understanding of the strain gradient effects, which will open up new technical opportunities for multifunctional device designs, such as high density data storage via mechanical means, nanoscale flexoelectric sensors, detectors, etc.

**Methods**

A NanoCute SPI 3800 (Hitachi HiTech Science) SPM system which enables both piezoresponse force microscopy (PFM) and magnetic force microscopy (MFM) measurements was used to investigate local piezoelectric/ferroelectric and magnetic properties at room temperature. A conduction tip (Si cantilever coated with Rh) was used to apply electric field and mechanical force in PFM measurements, while a magnetic tip coated with CoPtCr film was used in the MFM measurements. The reason for not using the MFM tip to do force scanning was to avoid the magnetization effect from the very weak magnetic field of the MFM tip. The mechanical constant was 15 N/m, and the resonance frequency was 139 kHz. A schematic representation of both PFM and MFM is presented in Figure 2(f).

The PFM mode is driven electrically using contact mode atomic force microscope (AFM) feedback for topographic tracing, and a periodic bias $V_{tip} = V_{dc} + V_{ac}cos(\omega t)$ is applied on the tip. The tip deflection is $A = A_0 + A_{1\omega}cos(\omega t + \theta)$, where the deflection amplitude $A_{1\omega}$ is determined by the tip motion, and the phase ($\theta$) indicates the orientation of the atomic polarization.[27] When the electric field is applied to the piezoelectric material, domains with an upward polarization vector contract with a negative voltage, producing a phase shift of $\theta = 0^{\circ}$. For the downward domains, the situation is reversed, and $\theta = 180^{\circ}$. The pizoresponse amplitude ($A = A_{1\omega}/V_{ac}$) defines the local electromechanical activity of the surface.[28] In our Nanocute SPM system, the PFM image is recorded as an *Acos* image which combines both amplitude and phase components, making it easier to observe both the polarization and the electromechanical response in the sample using our SPM. In this work, *Acos* images were recorded with topographic images, applying a signal at 5 kHz with a 2 V oscillation under a ±15 V dc bias. We didn't apply dc bias on the tip when using mechanical force to switch the sample. The force is applied only in VPFM mode.

Each MFM scan included two steps: firstly, the trace was recorded in tapping mode to image the topography of the surface; and the second step was in lift mode, in which the tip does not come into contact with the film surface, to assess the stray magnetic field perpendicular to the surface. The lift distance of the probe tip was about 10 nm during the MFM measurements.

In order to plot the PFM polarization switching as a function of loading force [Figure 3(d)], we measured the polarization switching by recording the response voltage difference ($\Delta mV$) between the electrically switched area and the mechanically switched area in the *Acos* image [inset of Figure 3(d)]. The corresponding magnetic switching as a function of loading force was also recorded in terms of the magnetic difference ($\Delta^{\circ}$) in the MFM phase image [inset of Figure 3(e)].

**Acknowledgements**

Part of this work was supported by a Green Network of Excellence (GRENE) project of the Ministry of Education, Culture, Sports, Science and Technology (MEXT) of Japan.

Author T.T. Jia is grateful for support by grants from the Japan Society for the Promotion of Science (JSPS) (No. 26-04206). Author Z. X. Cheng thanks the Australian Research Council (ARC) for support through a Future Fellowship. Author H.Y. Zhao thanks the National Natural Science Foundation of China for support (No. 51402327).

Correspondence and Requests for materials should be addressed to:

Dr. Hideo Kimura

National Institute for Materials Science, 1-2-1 Sengen, Tsukuba, Ibaraki 305-0047, Japan

E-mail:  kimura.hideo@nims.go.jp

A/Prof. Zhenxiang Cheng

Institute for Superconducting & Electronic Materials, University of Wollongong, Innovation Campus, North Wollongong, NSW 2500, Australia

E-mail: cheng@uow.edu.au

# References

1.      Eerenstein W, Mathur ND, Scott JF. Multiferroic and magnetoelectric materials. *Nature* **442**, 759-765 (2006).

2.      Vaz CAF, Hoffman J, Anh CH, Ramesh R. Magnetoelectric Coupling Effects in Multiferroic Complex Oxide Composite Structures. *Adv Mater* **22**, 2900-2918 (2010).

3.      Spaldin NA, Fiebig M. The Renaissance of Magnetoelectric Multiferroics. *Science* **309**, 391-392 (2005).

4.      Wang J, *et al.* Epitaxial BiFeO$_3$ multiferroic thin film heterostructures. *Science* **299**, 1719-1722 (2003).

5.      Chen YC, *et al.* Electrical Control of Multiferroic Orderings in Mixed-Phase BiFeO$_3$ Films. *Adv Mater* **24**, 3070-3075 (2012).

6.      Ederer C, Spaldin N. Weak ferromagnetism and magnetoelectric coupling in bismuth ferrite. *Phys Rev B* **71**, 060401 (2005).

7.      Kogan SM. Piezoelectric effect during inhomogeneous deformation and acoustic scattering of carriers in crystals. *Sov Phys Solid State* **5**, 2069-2070 ( 1964 ).

8.      Jeon BC, *et al.* Flexoelectric effect in the reversal of self-polarization and associated changes in the electronic functional properties of BiFeO$_3$ thin films. *Adv Mater* **25**, 5643-5649 (2013).

9.      Lu H, *et al.* Mechanical writing of ferroelectric polarization. *Science* **336**, 59-61 (2012).

10.    Lu H, *et al.* Mechanically-induced resistive switching in ferroelectric tunnel junctions. *Nano Lett* **12**, 6289-6292 (2012).

11.    Heo Y, Jang B-K, Kim SJ, Yang C-H, Seidel J. Nanoscale Mechanical Softening of Morphotropic BiFeO$_3$. *Adv Mater* **26**, 7568-7572 (2014).

12.    Gregg JM. Applied physics. Stressing ferroelectrics. *Science* **336**, 41-42 (2012).

13.    Ma W, Cross LE. Strain-gradient-induced electric polarization in lead zirconate titanate ceramics. *Appl Phys Lett* **82**, 3293-3295 (2003).

14.    Lee S, *et al.* Giant magneto-elastic coupling in multiferroic hexagonal manganites. *Nature* **451**, 805-808 (2008).

15.    Zavaliche F, *et al.* Electrically Assisted Magnetic Recording in Multiferroic Nanostructures. *Nano Lett* **7**, 1586-1590 (2007).

16.    Nelson CT, *et al.* Domain dynamics during ferroelectric switching. *Science* **334**, 968-971 (2011).

17.    Heron JT, Schlom DG, Ramesh R. Electric field control of magnetism using BiFeO$_3$-based heterostructures. *Appl Phys Rev* **1**, 021303 (2014).

18.    Chu YH, *et al.* Nanoscale domain control in multiferroic BiFeO$_3$ thin films. *Adv  Mater* **18**, 2307-2311 (2006).

19.    Balke N, Bdikin I, Kalinin SV, Kholkin AL. Electromechanical Imaging and Spectroscopy of Ferroelectric and Piezoelectric Materials: State of the Art and Prospects for the Future. *J Am Ceram Soc* **92**, 1629-1647 (2009).

20.  Chu YH, *et al.* Electric-field control of local ferromagnetism using a magnetoelectric multiferroic. *Nat Mater* **7**, 478-482 (2008).

21.  Baek SH, *et al.* Ferroelastic switching for nanoscale non-volatile magnetoelectric devices. *Nat Mater* **9**, 309-314 (2010).

22.  Zhao T, *et al.* Electrical control of antiferromagnetic domains in multiferroic BiFeO$_3$ films at room temperature. *Nat Mater* **5**, 823-829 (2006).

23.  Jaafar M, Iglesias-Freire O, Serrano-Ramon L, Ibarra MR, de Teresa JM, Asenjo A. Distinguishing magnetic and electrostatic interactions by a Kelvin probe force microscopy-magnetic force microscopy combination. *Beilstein J Nanotech* **2**, 552-560 (2011).

24.  Zubko P, Catalan G, Tagantsev AK. Flexoelectric Effect in Solids. *Annu Rev Mater Res* **43**, 387-421 (2013).

25.  Gu Y, Hong Z, Britson J, Chen L-Q. Nanoscale mechanical switching of ferroelectric polarization via flexoelectricity. *App Phys Lett* **106**, 022904 (2015).

26.  Gich M, *et al.* Multiferroic Iron Oxide Thin Films at Room Temperature. *Adv Mater* **26**, 4645-4652 (2014).

27.  Kalinin SV, *et al.* Vector Piezoresponse Force Microscopy. *Microsc Microanal* **12**, 206-220 (2006).

28.  Kalinin SV, Shao R, Bonnell DA. Local Phenomena in Oxides by Advanced Scanning Probe Microscopy. *J Am Chem Soc* **88**, 1077-1098 (2005).

**Author contribution statement**

T.T.J., H. K. and Z.X. C. proposed the research project. T.T.J. prepared the thin films and carried out the characterizations. T.T.J, H.K. and Z.X.C. wrote the manuscript. All authors discussed the results and contributed to the manuscript.

**Figure Captions**

**Figure 1**. Interaction among polarization, magnetization, and the strain gradient in multiferroic thin films under external stimuli: magnetic field ($H$), electric field ($E$), and stress ($\sigma$).

**Figure 2**. PFM and MFM images of of $Bi_{0.9}La_{0.1}FeO_3$ thin film: (a) Topographic image, (b) VPFM and (c) LPFM box-in-box lithographic images poled by ±15 V, and (d) corresponding MFM image scanned in the same area as VPFM. (e) Schematic illustrations of the three kinds of domain walls permitted in $BiFeO_3$ for in-plane and out-of-plane orientations. (f) Schematic illustrations of PFM and MFM operations.

**Figure 3**. VPFM and MFM images of $Bi_{0.9}La_{0.1}FeO_3$ thin film: (a) Topographic image, (b) VPFM *Acos* image and (c) MFM phase image of the film poled by alternate dc bias of ±15 V and 800 nN. (d) PFM response difference ($\Delta mV$) of *Acos* image between the electrically switched area and the force switched area as a function of the loading force obtained in the Nanocute SPM system, (e) MFM phase difference ($\Delta^o$) between the electrically switched area and the mechanically switched area. Insets to (d) and (e) show the response voltage difference and magnetic phase difference measurements, respectively.

**Figure 4.** Vertical piezoresponse force microscope (VPFM) images of $Bi_{0.9}La_{0.1}FeO_3$ thin film: (a) Lithographic pattern created by alternate dc bias of ±15 V and external

mechanical force of 800 nN; (b) short-circuiting of the bottom electrode and the cantilever during pattern reading; (c) PFM image scanned after removal of the short-circuit. (d) Schematic diagram of the measurement patterns corresponding to the polarization directions.

1  **Figures**

2  Figure 1

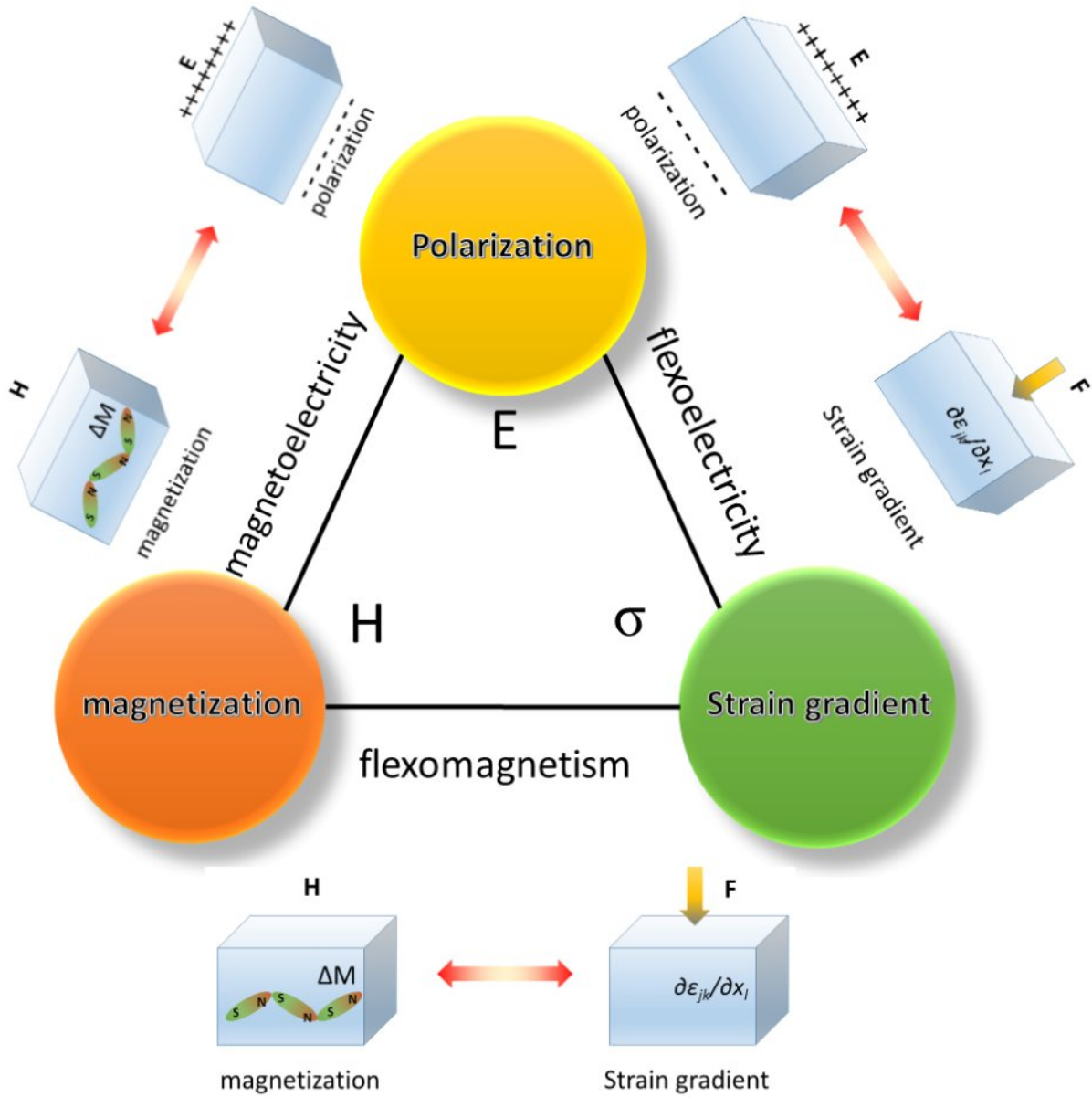

Switching of ferroelectric and magnetic domains by force

1  Figure 2

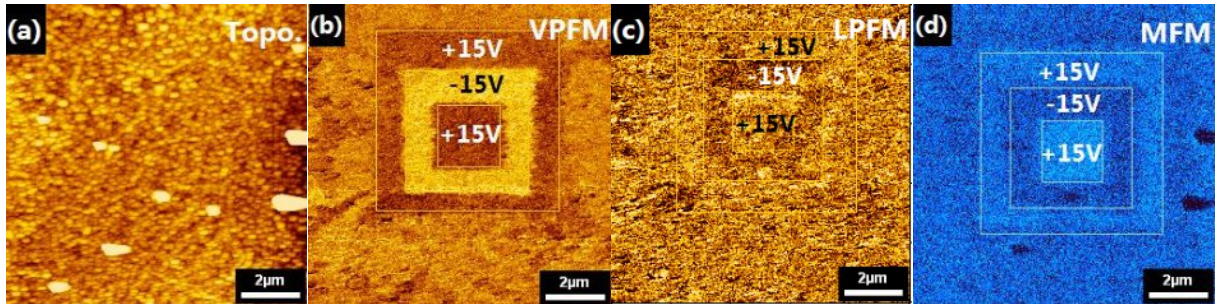

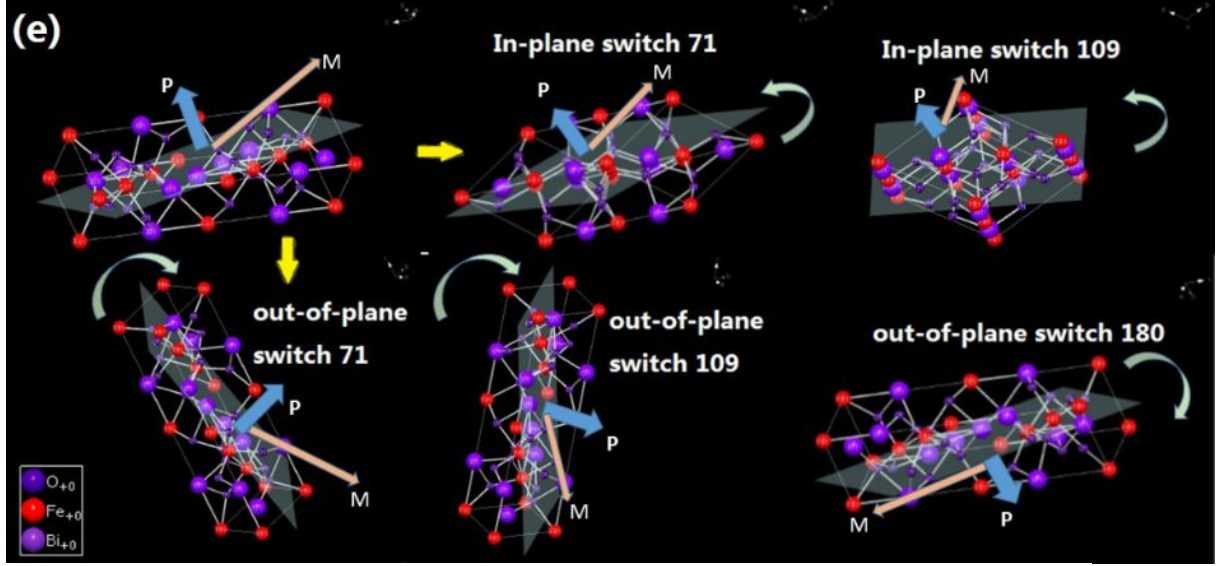

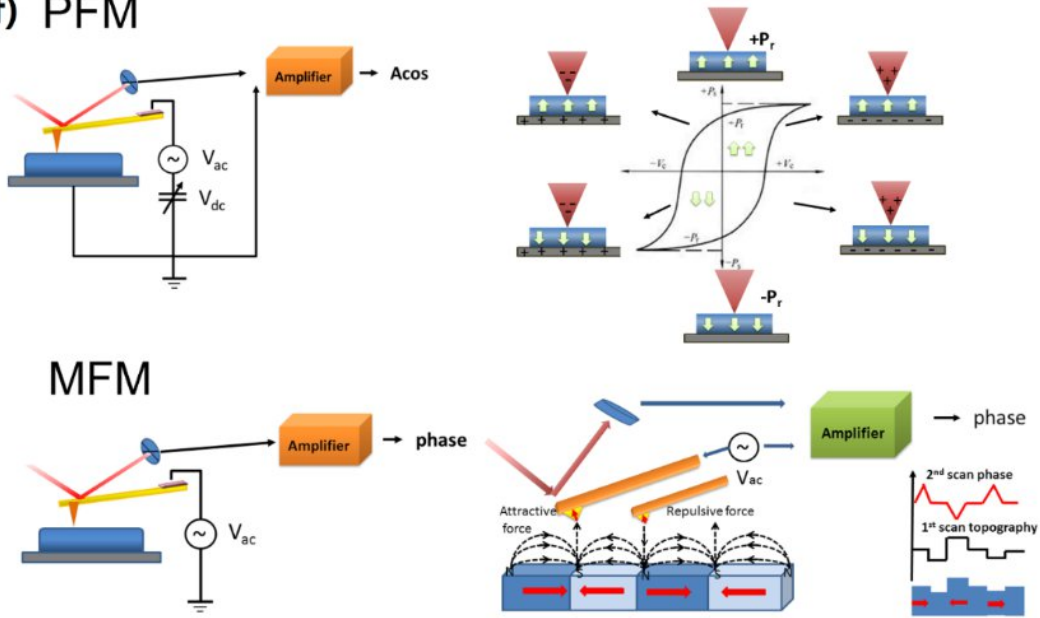

Switching of ferroelectric and magnetic domains by force

1    Figure 3

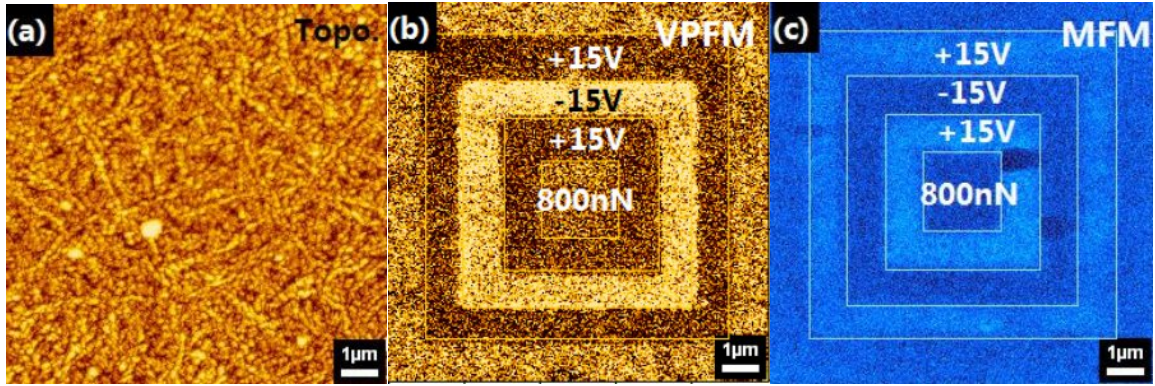

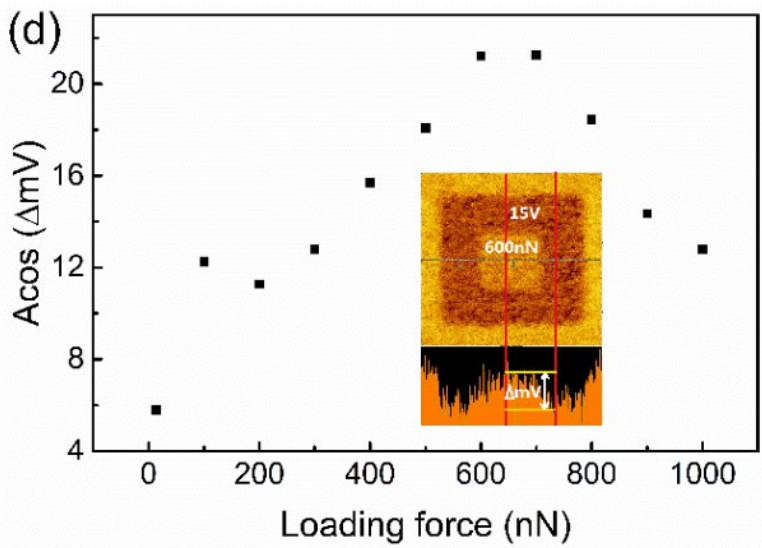

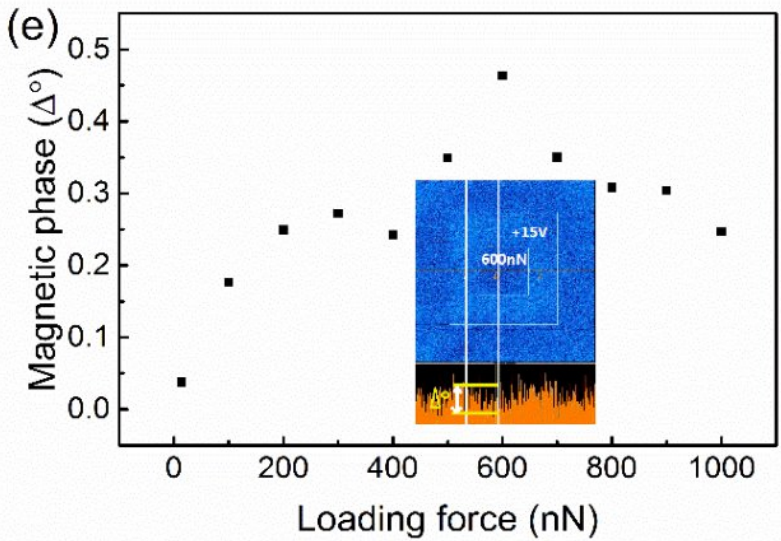

Switching of ferroelectric and magnetic domains by force

1     Figure 4

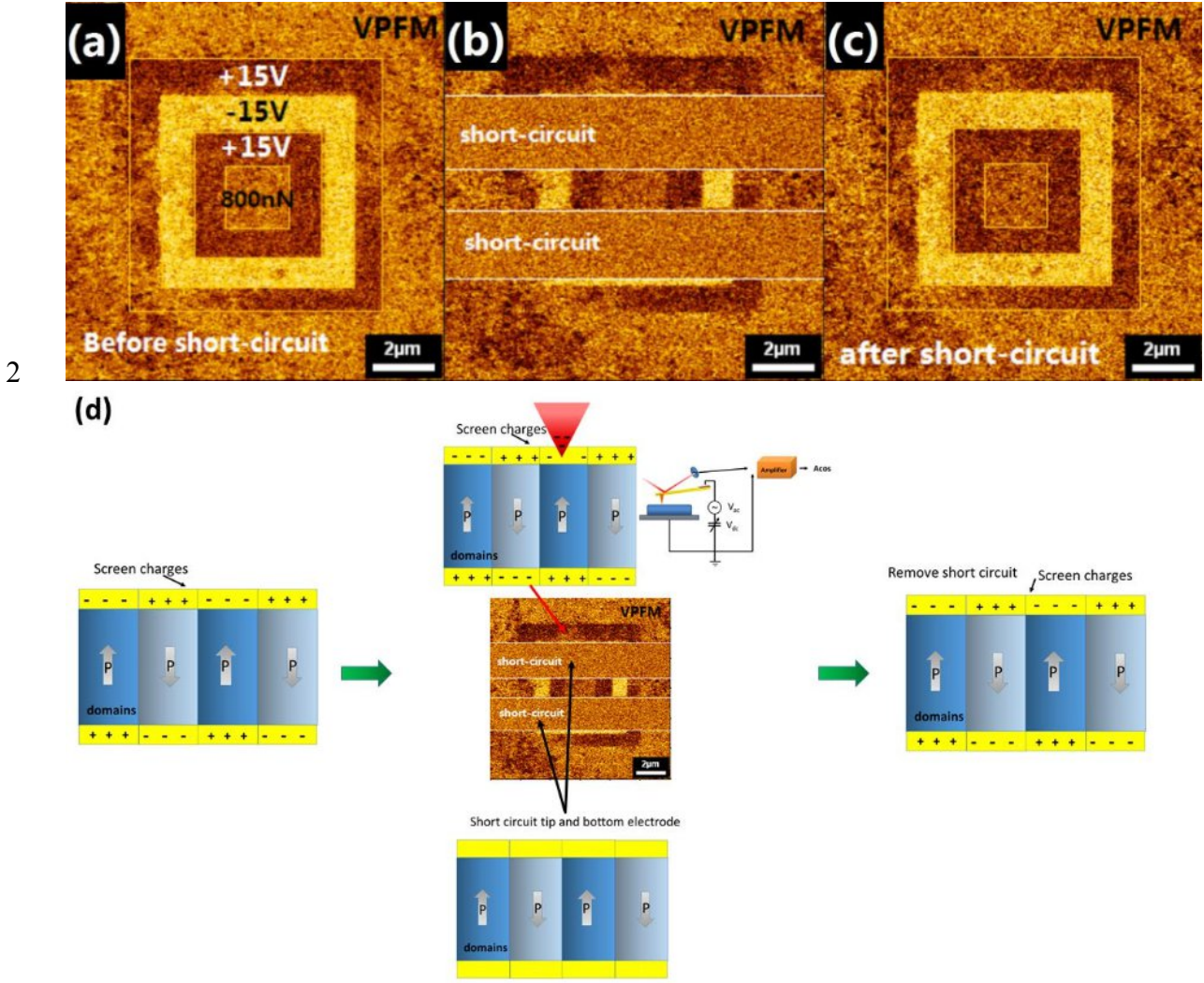

4     Supporting Information

8     **Switching of both local ferroelectric and magnetic domains in multiferroic Bi$_{0.9}$La$_{0.1}$FeO$_3$**

9     **thin film by mechanical force**

    Switching of ferroelectric and magnetic domains by force

Tingting Jia, Hideo Kimura[*], Zhenxiang Cheng[*], and Hongyang Zhao

Dr. T. T. Jia, Dr. H. Kimura

National Institute for Materials Science, 1-2-1 Sengen, Tsukuba, Ibaraki 305-0047, Japan

E-mail: kimura.hideo@nims.go.jp

Dr. Z. X. Cheng

Institute for Superconducting & Electronic Materials, University of Wollongong, Innovation

Campus, North Wollongong, NSW 2522, Australia

E-mail: cheng@uow.edu.au

Dr. H. Y. Zhao

Department of Materials Science and Engineering, Wuhan Institute of Technology, Wuhan

430074, China

Keywords: Scanning probe microscopy, bismuth ferrite, flexoelectric, mechanical force, magnetic domain.

**Film deposition.**

$Bi_{0.9}La_{0.1}FeO_3$ (BLFO) thin film was deposited on $Pt/TiO_2/SiO_2/Si$ substrates using a pulsed laser deposition (PLD) system with the laser source at 355 nm and a repetition rate of 10 Hz. The ceramic target for BLFO deposition was prepared using a conventional solid-state reaction process. The BLFO thin film was deposited at 550 °C over a period of 30 min, and the sample was then cooled down to room temperature (RT). The detailed deposition conditions can be found in the previous reports of our group.[1, 2]

Switching of ferroelectric and magnetic domains by force

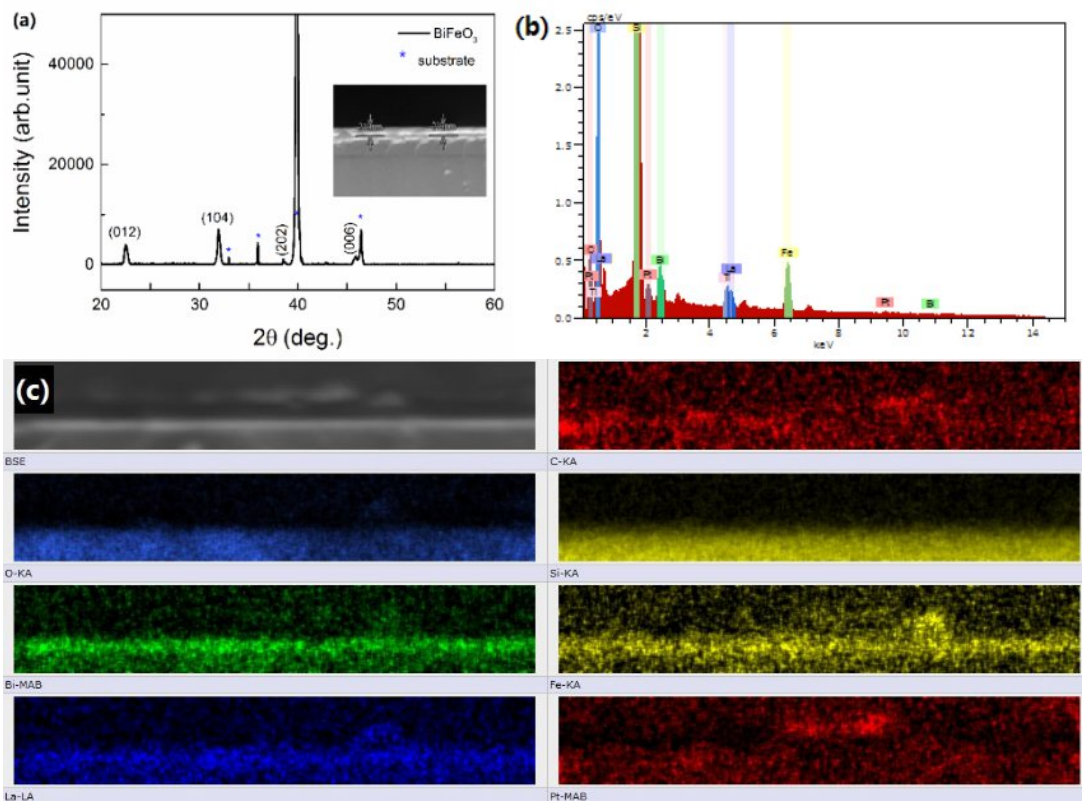

**Figure S1.** (a) XRD pattern of $Bi_{0.9}La_{0.1}FeO_3$ thin film deposited on $Pt/TiO_2/SiO_2/Si$ substrate, with the inset showing a cross-sectional SEM image of the as-deposited thin film. (b) EDX element analysis spectrum, and (c) element mapping.

The crystal structure was evaluated by X-ray diffraction (XRD) using a diffractometer (RIGAKU RINT 2000). The film thickness was measured using a JSM-6500F scanning electron microscope (SEM). **Figure S1**. shows the XRD pattern of the as-deposited BLFO thin film deposited on $Pt/TiO_2/SiO_2/Si$ substrate. The film was found to be polycrystalline and based on a hexagonal symmetry with space group R3c (161), with lattice parameters of $a = 0.5746$ nm, $b = 0.5746$ nm, and $c = 1.3410$ nm. The as-deposited film shows evidence of a random growth habit on the $Pt/TiO_2/SiO_2/Si$ substrate with a mixture of (012), (104), (202), and (006) diffraction peaks. The

Switching of ferroelectric and magnetic domains by force

1  cross-sectional SEM image in the inset of Figure S1(a) shows the clear structure of the as-

2  deposited BLFO film deposited on $Pt/TiO_2/SiO_2/Si$ substrate with film thickness of ~219 nm. At

3  this thickness, a piezoresponse can be recorded easily, and the minimum switching field

4  increases when the film thickness is increased. The energy dispersive X-ray spectroscopy (EDX)

5  element analysis shown in Figure S1(b) qualitatively indicates that the main components in the

6  as-deposited BLFO thin film are Bi, La, Fe, and O, which matches well with the chemical

7  composition of the film. The element mapping [Figure S1(c)] over the entire film shows a

8  uniform distribution of the main elements.

Switching of ferroelectric and magnetic domains by force

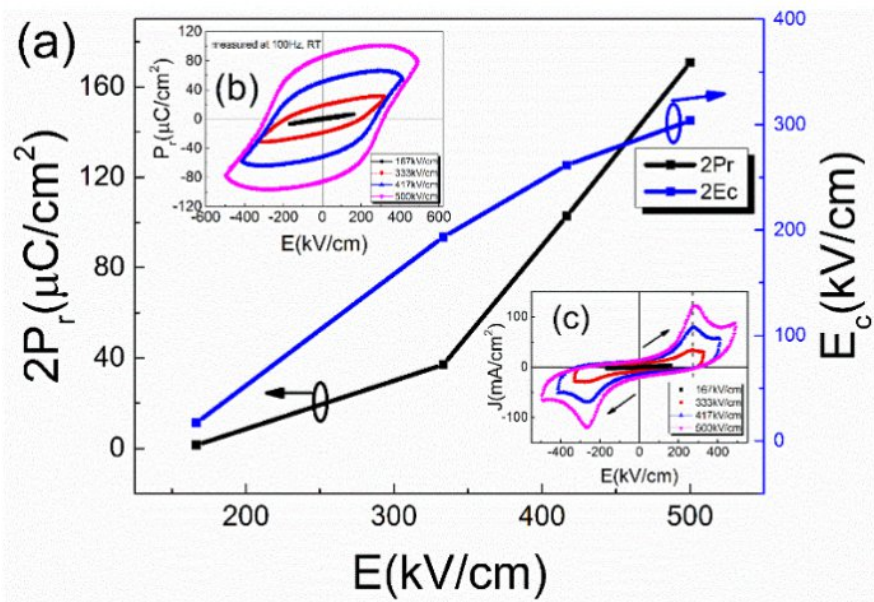

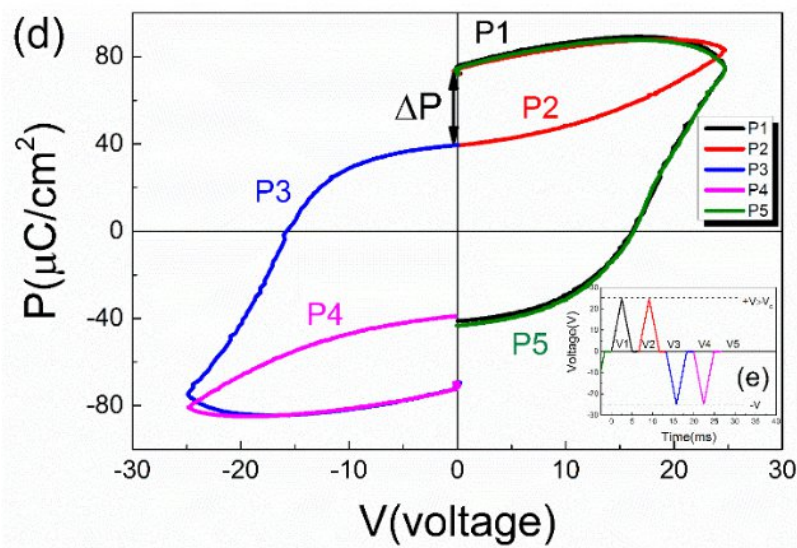

4  **Figure S2.** Ferroelectric measurements of Pt/BLFO/Pt capacitor: (a) remanent polarization and

5  coercive field as a function of electric field; (b) ferroelectric polarization − electrical field (*P-E*)

6  hysteresis loops measured at different electric fields; (c) leakage current − electric field curves

7  during *P-E* loop measurements; (d) positive-up negative-down (PUND) switching polarization as

Switching of ferroelectric and magnetic domains by force

1  a function of voltage; (e) waveform of the applied triangle pulse with a 1 ms read pulse time and

2  1 ms delay.

4  For electrical measurements, a Pt top electrode was coated on the surface of the BLFO thin film

5  through a shadow mask with a diameter of 100 μm to form a capacitor. The ferroelectric

6  properties were measured at room temperature (RT) with an aixACCT TF-1000 ferroelectric

7  tester. Figure S2(a) shows the remanent polarization and coercive field of the Pt/BLFO/Pt

8  capacitor as a function of applied electric field. A large remanent polarization ($2P_r$) of 170.7

9  $\mu C/cm^2$ and a coercive field ($E_c$) of about 303.6 kV/cm at a maximum applied field of 500 kV/cm

10 were observed. The ferroelectric hysteresis ($P$-$E$) loops measured at different applied fields are

11 shown in Figure S2(b). Figure S2(c) shows leakage current − electric field curves ($J$-$E$) during $P$-

12 $E$ loop measurements. Obviously, there are current density peaks corresponding to the switching

13 of polarization by an external field. In addition, a linear current contribution can be identified,

14 which suggests that the polarization of BLFO in the $P$-$E$ loops contains some contribution from

15 the leakage current. To exclude the contribution of the leakage current to the polarization and to

16 identify the intrinsic polarization value of our film, we carried out so-called positive-up-

17 negative-down (PUND) measurements. Figure S2(d) shows the switching polarization as a

18 function of the applied voltage, and the inset Figure S2(e) shows the applied voltage waveform.

19 As shown in Figure S2(d), the switched polarization is saturated with the switching time and is

20 consistent with the $2P_r$ value obtained from the hysteresis measurement. The pulsed remanent

21 polarization value $\Delta P_r$ [switched polarization ($P_{sw}$) - non-switched polarization ($P_{nsw}$)] (82.7

22 $\mu C/cm^2$) is approximately 1.2 times that of $P_r$ (70.6 $\mu C/cm^2$). Thus, the presence of intrinsic

23 ferroelectricity in our BLFO thin film can be concluded.

Switching of ferroelectric and magnetic domains by force

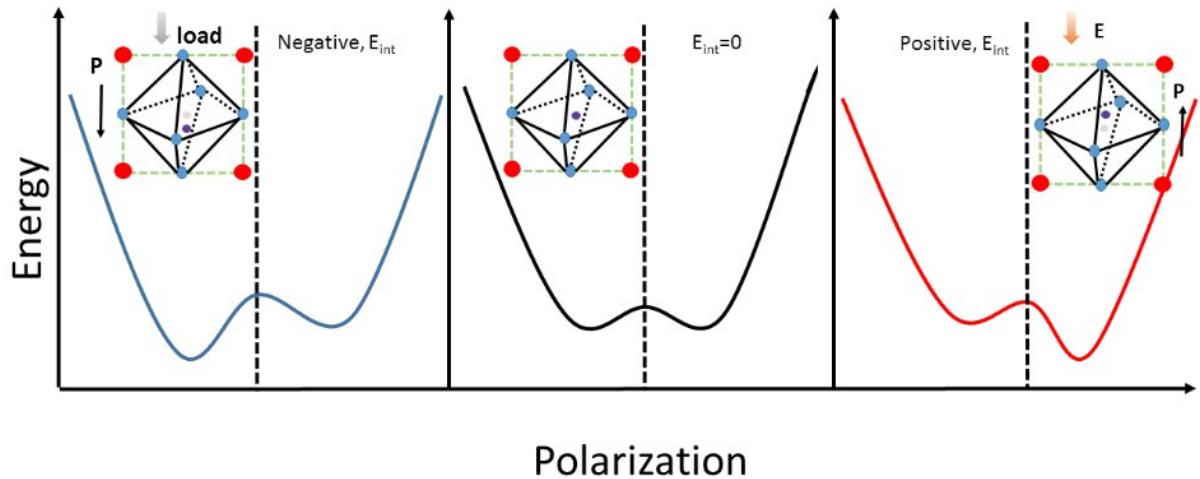

**Figure S3.** Schematic illustration of the relationship between free energy and polarization. The free energy of the BLFO film: (a) with load, (b) without any extra loading force or electric field,

(c) with positive electric field.

As illustrated in Figure S3, the Gibbs free energy of the thin film is symmetric in the equilibrium

state. The loading force of the tip induces an inhomogeneous strain gradient in the system, giving

rise to a flexoelectric field across the film, which changes the free energy profile asymmetrically:

it destabilizes the positive side of the double well and forces downward switching on the negative side. The electric field with a positive direction, in contrast, modifies the free energy asymmetrically on the positive side.

Switching of ferroelectric and magnetic domains by force

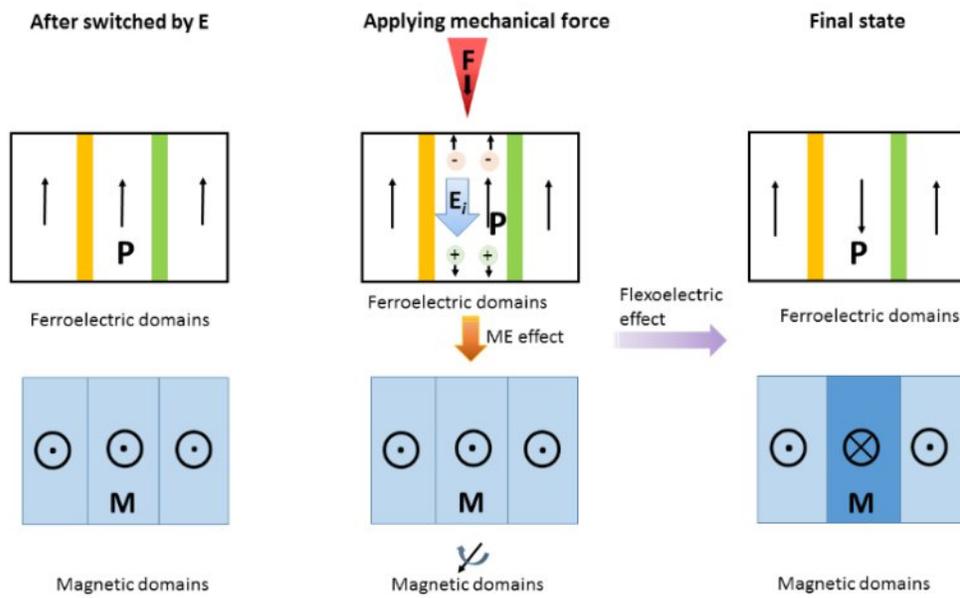

2 **Figure S4.** Illustration of ferroelectric domain switching and magnetic domain switching when

3 applying mechanical force.

5 The ferroelectric domains and magnetic domains are initially switched by an electric field, and

6 the polarization of the ferroelectric domains and the antiferromagnetic orientation of the

7 magnetic domains are the same. On applying mechanical force on the surface, an internal electric

8 field ($E_i$) is generated due to the flexoelectric effect, which leads to charge separation and a

9 resulting polarization reversal in the ferroelectric domains, so that magnetic domain switching

10 occurs via the ME coupling effect in the BLFO thin film. The direction of the spatial rotation of

11 the magnetization follows a cycloid-like path. Finally, both the ferroelectric domains and the

12 magnetic domains are switched by mechanical force.

13 References

14 1.      Cheng ZX, Wang XL, Kimura H, Ozawa K, Dou SX. La and Nb codoped BiFeO$_3$

15 multiferroic thin films on LaNiO$_3$/Si and IrO$_2$/Si substrates. *Appl Phys Lett* **92**, 092902 (2008).

Switching of ferroelectric and magnetic domains by force

1    2.    Cheng ZX, Wang XL, Dou SX, Kimura H, Ozawa K. Improved ferroelectric properties in

2    multiferroic $BiFeO_3$ thin films through La and Nb codoping. *Phys Rev B* **77**, 092101 (2008).

Switching of ferroelectric and magnetic domains by force